# Interactive MCQs as a tool for Knowledge Acquisition

[1]Kisor Ray

*Techno India Agartala,Tripura University*

ray.kisor@gmail.com

[2]Saumen Sarkar

*NSEC (Techno India Group),West Bengal University of Technology*

*saumensarkar@hotmail.com*

**Abstract :** Multiple Choice Questions or MCQs are very important for e-learning. Generally, MCQs are used as a tool for the assessment of students' performance at the end of their learning sessions. Can MCQs become an important tool in the process of knowledge acquisition while attending a course? This paper intends to find out how MCQs could be used as a tool for the better understanding, coverage as well as knowledge acquisition.

## I. INTRODUCTION

MCQs are used very widely across the globe to evaluate students' performance by the various academic institutes and universities starting from the entrance examinations to periodic tutorials, assignments and semester examinations. Corporate houses and public sectors use MCQs for the screening of the prospective employees. These are used for general quiz competitions organized by different entities including the electronic ones. MCQs are widely used by the e-learning portals for quizzing the participants after the end of the courseware to find the level of knowledge being gathered by them. E-learning needs a couple of components, MCQs being one of them, to serve as a knowledge evaluation tool. E-learning offers online courseware and computer assisted assessment thus reducing the burden of managing a large pool of students compared to traditional paper based testing process. While MCQs are viewed as a tool for the assessment of learning, this paper examines whether this could be used as a very effective knowledge acquisition tool[1].

## II. MOTIVATION

Interest of the students is the key for the success of e-learning courseware. The greater the interest, more involvement and more understanding about the subjects. The success or failure of a courseware is related to student motivation. An interesting courseware is more likely to generate better motivation. A better motivated student is more likely to put better effort thus producing better performances. So, each of the important components of an e-learning courseware needs to be examined carefully at the implementation level. This paper intends to examine the varied implementation of different MCQs as well as different feedback types and their effects on the performance level of the participating students in order to suggest certain implementation styles while designing an e-learning courseware for the quizzing purpose.

## III. METHODS AND PROCEDURES

In the process of our investigation, we have used 7 different types of MCQs as given below:

1. True/False
2. Multiple Choice
3. Multiple Response
4. Fill in the Blank
5. Matching
6. Numeric/Number Range
7. Hotspot

Among these 7 different types, we have used GIFT formatting to generate type 1 to type 6. GIFT format allows one to use a text editor to write quiz questions in a simple format that can be imported into Adobe Captivate or Moodle Quiz. This makes it easier to port the quizzes by just producing them using simple text editors like Notepad, Notepad++ under Windows or nano and vi under Linux/Unix operating system. The quizzes were created in three categories: (I) Quiz with detailed feedback (QC1) (II) Quiz with short feedback (QC2) (III) Quiz without feedback (QC3). These quizzes were used under two different situations: (1) Courseware Practice Test (CPT) - untimed (2) Courseware Evaluation Test (CET) – timed. We used all the three categories of quizzes for the courseware practice test purpose but for the evaluation test quiz with no feedback only. Three different subjects namely : Engineering Physics, Engineering Chemistry and Basic

Electrical were chosen following the syllabus for the students of 1st semester of our Institute, to create 9 different quizzes under three categories.

| Subject Name | Quiz Code | Quiz Type |
|---|---|---|
| Basic Electrical | QBE1 | Quiz with detailed feedback (QC1) |
| Basic Electrical | QBE2 | Quiz with short  feedback(QC2) |
| Basic Electrical | QBE3 | Quiz with no  feedback(QC3) |
| Engineering Physics-I | QEP1 | Quiz with detailed feedback (QC1) |
| Engineering Physics-I | QEP2 | Quiz with short  feedback(QC2) |
| Engineering Physics-I | QEP3 | Quiz without   feedback(QC3) |
| Engineering Chemistry-I | QEC1 | Quiz without feedback (QC1) |
| Engineering Chemistry-I | QEC2 | Quiz with short  feedback(QC2) |
| Engineering Chemistry-I | QEC3 | Quiz without feedback(QC3) |

Table 1. Nine different quizzes under three categories

Among these three categories of quizzes, category 1 is the quiz with detailed feedback. While a student attempts a question, irrespective of the 'right' or 'wrong' result a detailed explanation about the correct answer appears on the screen. Category 2 is the quiz with a short feedback. While students attempt a question, a short explanation appears on the screen. Category 3 is the quiz without any feedback. While a student attempts a question, no explanation appears or pops up except the right or wrong notifications.

Example of a Category 1 Quiz (QC1) with detailed feedback

*Kepler's second law regarding constancy of arial velocity of a planet is a consequence of the law of conservation of*

(a) force
(b) angular momentum
(c) linear momentum
(d) energy

Answer : (b) An imaginary line drawn from the center of the sun to the center of the planet will sweep out equal areas in equal intervals of time'. The ellipse traced by a planet around the Sun has a symmetric shape, but the **motion** is not symmetric. This means that the planet moves faster when it is near the sun, slower when it is far away. Kepler's second law of planetary motion is a direct consequence of angular momentum conservation.

Example of  a Category 2 Quiz (QC2) with short feedback
*Kepler's second law regarding constancy of arial velocity of a planet is a consequence of the law of conservation of*

(a) force
(b) angular momentum
(c) linear momentum
(d) energy

Answer : (b) Kepler's second law of planetary motion is a direct consequence of angular momentum conservation.

Example of a Category 3 Quiz (QC3) without feedback
*Kepler's second law regarding constancy of arial velocity of a planet is a consequence of the law of conservation of*
(a) force
(b) angular momentum
(c) linear momentum
(d) energy
Answer : (b)

Each of the quizzes of different categories were created on three different topics of each subject following the guidelines : (1) Relatively harder topic will have the QC1 quiz type (2) Relatively medium hard topic will have the QC2 quiz type (3) Relatively easy topic will have QC3 quiz type. At the of the completion of courseware on the three subjects: Engineering Physics-1 (EP-I), Engineering Chemistry (EC-I) and Basic Electrical (BE), students had to take Courseware Practice Test (CPT) on each topic. The CPT was mandatory but without any time constraint. Students then had to take Courseware Evaluation Test (CET) for each subject which is a timed test .CET was created from the pool of the existing question banks of type QC1,QC2 and QC3 along with new additional conceptual questions. The CET was designed to end beyond the specified time and students' score for each subject to get recorded at that point.118 students of 1st semester of our Institute from different engineering branches like Civil, Mechanical, Electrical, Electronics and Computer Science  had to go through the process of studying the courseware, practice tests and evaluation tests on the subjects mentioned earlier. Their performances were recorded and tabulated for the further study and investigation.

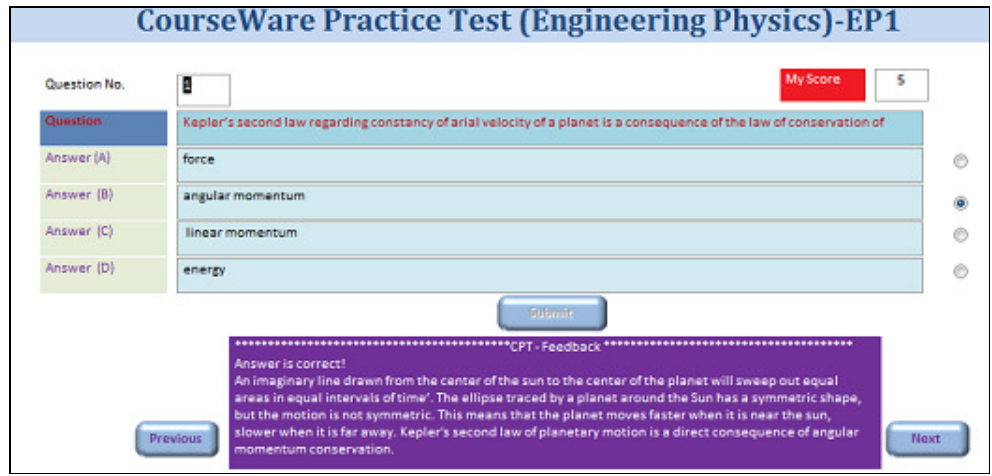

Fig. 1  A sample detailed feedback (locked screen) for a correct answer

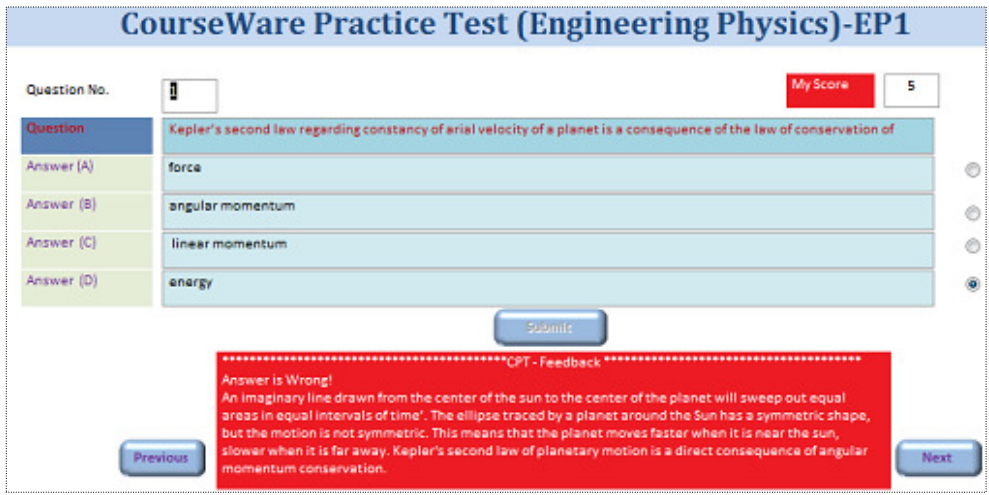

Fig. 2  A sample detailed feedback (locked screen) for a wrong answer

IV. **FINDINGS**

| Subject | Quiz Code | Topic | Quiz Type used in CPT | Average Score in CET |
|---------|-----------|-------|-----------------------|----------------------|
| Basic Electrical | QBE1 | BET1 | QC1 | 76% |
| Basic Electrical | QBE2 | BET2 | QC2 | 68% |
| Basic Electrical | QBE3 | BET3 | QC3 | 61% |
| Engineering Physics-I | QEP1 | EP-IT1 | QC1 | 74% |
| Engineering Physics-I | QEP2 | EP-IT2 | QC2 | 66% |
| Engineering Physics-I | QEP3 | EP-IT3 | QC3 | 59% |
| Engineering Chemistry-I | QEC1 | EC-IT1 | QC1 | 72% |
| Engineering Chemistry-I | QEC2 | EC-IT2 | QC2 | 62% |
| Engineering Chemistry-I | QEC3 | EC-IT3 | QC3 | 58% |

Table 2: Students' performance score in the Evaluation Test

Table 2 shows practices tests based on QC1 types of feedback (detailed answer as mandatory review) could generate better knowledge gathering for the participating students for all the subjects as a result they could perform better in the respective subjects in the courseware evaluation tests (CET).This pattern is uniform. Though at this point of time we did not include any graphical feature and/or animation for the feedback purpose other than plain and simple texts ,it is understood that should we provide interacting animated feedback during the courseware practice tests without any time restriction, students are most likely to find it more interesting as well as a handy tool for the knowledge acquisition purpose.

## V. CONCLUSION

What we found that if we insert a detailed explanation of a question's answer and display it  on the screen for a reasonable duration as a mandatory display (by locking the screen for a short duration) which cannot be bypassed and/or escaped, students are forced to review the correct and detailed answers  irrespective of their right/wrong answers (which also eliminates the chance of missing a right answer even though the student scored without proper knowledge and/or using guessing technique).Should we prepare e-learning courseware with detailed feedback  about the correct answer at the Courseware Practice Test(CPT) level and make it technically mandatory for the students to review, they are most likely to perform better in that particular subject and/or course. Most of the modern quizzing tools including but not limited to Adobe Captivate, Moodle, Articulate Storyline, HotPotatos, ClassMarker, Classtools, Quizlet etc. are having some feedback mechanisms for the questions created by these tools. Detailed answers may be inserted into the quizzes for each questions at practice test level. Most of these tools by default do not offer any screen-locking mechanism for forcing the students to read the 'detailed answer' explained on the screen. However, a good numbers of them alternatively offer implementation of the same by creating special scripts or triggers for which some programming knowledge will be mandatory. A recommended good practice would be to record the results of the Courseware Practice Test(CPT),Courseware Evaluation Test(CET) and the Subjects as well as the percentage of  similar types and/or same set of questions which were used as a part of CPT with QC1 type and were repeated at CET. The same could be repeated for type QC2 and QC3. The result would most likely indicate the QC1 type would produce a better result over the other types. A nicely designed database (for which MS Access, PostgreSQL , MySQL and/or any other similar database) could be used to capture and store all the results in a systematic way for the purpose of periodic evaluation[2][3]. Without any complicated programming, using MS Excel with appropriate ODBC drivers, good reports and charts could be produced for the further study. Our investigation does not include any exclusive finding on the type of MCQs (e.g. True/False, Multiple Choice, Multiple Response, Matching, Numeric/Number Range, Hotsopt etc.) more effective with the QC1 type since we have used a mix and match of all the types of MCQs without tracking the types.

.